\documentclass[a4paper,11pt]{article}
\usepackage{jheppub}
\usepackage{color}
\newtheorem{conjecture}{Conjecture}[section]
\newtheorem{definition}{Definition}[section]
\newtheorem{theorem}{Theroem}[section]

\def\beq#1\eeq{\begin{align}#1\end{align}}

\title{K stability and stability of chiral ring}

\author[a]{Tristan C. Collins}
\author[b,c]{Dan Xie}
\author[a,b,c]{Shing-Tung Yau}

\affiliation[a]{Department of Mathematics, Harvard University, Cambridge, MA 02138, USA}
\affiliation[b]{Center of Mathematical Sciences and Applications, Harvard University, Cambridge, 02138, USA}
\affiliation[c]{Jefferson Physical Laboratory, Harvard University, Cambridge, MA 02138, USA}

\abstract{We define a notion of stability for chiral ring of four dimensional $\mathcal{N}=1$ theory by introducing test chiral rings and generalized a maximization. We conjecture that a chiral ring is the chiral ring of a superconformal field theory if and only if it is stable. 
 We then study $\mathcal{N}=1$ field theory derived from D3 branes probing a 
three-fold singularity $X$, and show that the K stability which implies the existence of Ricci-flat conic metric on $X$ is equivalent to the stability of chiral ring of the corresponding field theory. }

\begin{document} 
\maketitle
\flushbottom
%%%%%%%%%%%%%%%%%%%%%%%%%%%%%%%%%%%%%%%%%%%%%%%%%%%%%%%%%%%

\section{Introduction}
 The chiral ring of a four dimensional $\mathcal{N}=1$ theory plays a crucial role in understanding the dynamics of the theory.  In particular, the chiral ring can be used to determine the structure of the moduli space of vacua and the phase structure \cite{Cachazo:2002ry}.  Moreover, the chiral ring structure seems to be still quite important even if the theory has 
a unique vacua \cite{Xie:2016hny,Buican:2016hnq}.

However, little is known about the  general structure of the chiral ring of  $\mathcal{N}=1$ theory.  The purpose of this 
paper is to study the chiral ring of a superconformal field theory (SCFT).  We ask the following question:  when is a chiral ring ${\cal R}$
the chiral ring of a SCFT?  An obvious necessary condition is that the chiral ring has to be graded, since for a SCFT there is 
always a $U(1)_R$ symmetry which acts non-trivially on all of the chiral operators.  In particular, we really should start with a polarized chiral ring $({\cal R}, \zeta)$ and ask whether it is 
a chiral ring of a SCFT with $U(1)_R$ symmetry $\zeta$.  On the other hand, It is also known that the existence of a grading is not sufficient.

A second motivation for asking above question is the following: in many studies of supersymmetric field theory we start with an asymptotically free gauge theory ${\cal T}$ and assume that 
it flows to a SCFT ${\cal T}_0$ at a certain point of the moduli space (often the most singular point).  We can compute the chiral ring ${\cal R}$ of the theory ${\cal T}$, and let us denote by ${\cal R}_{0}$ the chiral ring of ${\cal T}_0$.  Many interesting quantities of the SCFT ${\cal T}_0$ can be computed if ${\cal R}= {\cal R}_0$.  For example, we can use $a$ maximization to determine the $U(1)_R$ symmetry \cite{Intriligator:2003jj} of ${\cal T}_0$. In general, however, ${\cal R}_0$ can differ from ${\cal R}$, for example:
\begin{itemize}
\item[(a)] It is believed that the chiral ring ${\cal R}$ of $\mathcal{N}=1$ $SU(N_c)$ SQCD with ${3\over 2}N_c<N_f<3N_c$ is the chiral ring ${\cal R}_0$ of 
the SCFT at the origin of the moduli space \cite{Seiberg:1994bz,Intriligator:1995au};
\item[(b)] If $ N_c<N_f\leq {3\over 2}N_c$, the chiral ring ${\cal R}$ of SQCD is not the chiral ring ${\cal R}_0$ of the SCFT at the origin, as the mesons 
become free at the SCFT point \cite{Seiberg:1994bz,Intriligator:1995au}; 
\item[(c)] A trivial example is a chiral scalar $\phi$ with cubic superpotential, and the chiral ring ${\cal R}$ of this theory is generated by the ideal $\phi^2=0$. However, the superpotential 
is marginally irrelevant at SCFT point, and the IR SCFT is free so its chiral ring ${\cal R}_0$ is free generated by operator $\phi$ and is different from ${\cal R}$.
\end{itemize}
From above examples, we learn that the possible reasons for ${\cal R}$ failing to be the chiral ring of a SCFT are: 
\begin{itemize}
\item Some operators hit the unitarity bound and become free at SCFT point, and this also modifies the chiral ring.
\item  Certain superpotential term is irrelevant at the SCFT point, and we should not impose the constraint from superpotential  for chiral operators of SCFT ${\cal T}_0$ \footnote{Notice 
that we can not ignore such superpotential term for ${\cal T}$ as it could be relevant at other vacua, and they are called dangerously irrelevant operator in \cite{Kutasov:1995ss}.}; 
\item There might be some other unknown dynamics that would lead to different chiral ring for SCFT.  We do not have a systematical way to detect them.
\end{itemize}
We can learn several interesting lessons from the above examples. Firstly ${\cal R}_0$ has more symmetries than ${\cal R}$.  Namely, there is a new symmetry generator acting on ${\cal O}$ alone if  ${\cal O}$ hits unitarity bound and becomes free; If a superpotential term formed by an  operator ${\cal O}$ becomes irrelevant, there could be a new symmetry acting on this operator ${\cal O}$ alone.
Secondly, ${\cal R}_0$ either leads to higher central charge $a$ \footnote{This does not violate the $a$ theorem, as ${\cal R}$ is not the chiral ring of a SCFT.},  or the same central charge as evidenced by example (c), but no less central charge.

Motivated by above examples, we introduce a notion of stability on chiral ring to characterize whether ${\cal R}={\cal R}_0$ and 
this notion also gives a method to define the chiral ring of a SCFT. The definition involves two basic elements: test chiral ring and generalized a maximization. 

Let's first discuss the test chiral ring. From above examples, if the chiral ring ${\cal R}$ fails to be the chiral ring of a SCFT, there is an associated different chiral ring ${\cal R}_0$: ${\cal R}_0$ can be derived by forgetting some of the superpotential terms if  ${\cal R}$ is derived from a quiver gauge theory, etc.  More generally ${\cal R}_0$ 
should have more symmetries, and it should satisfy certain continuity condition with respect to ${\cal R}$. Based on those observations, we propose:
\begin{definition}
A test chiral ring ${\cal R}_0$ can be derived from ${\cal R}$ by using a symmetry generator $\eta$ on  ${\cal R}$ and taking a \textbf{flat} limit. 
\label{def1}
\end{definition}
 Let's discuss more precisely what this definition means. Assume that  the chiral ring ${\cal R}$ is given by 
\begin{equation}
R={C[x_0,x_1,\ldots, x_n]\over I},
\end{equation}
here $x_i,i=0,\ldots,n$ are the generators of chiral ring and $I=(f_1,f_2,\ldots, f_m)$ is the ideal which gives the chiral ring relation among the generators. Now 
consider a one parameter subgroup $\eta(t)$ of $C^{n+1}$ and define its action on the  elements of idea $I$ as 
\begin{equation}
f(t)=\lambda(t)\cdot f=f(\lambda(t)\cdot(x_0,x_1,\ldots, x_n)).
\end{equation}
 So we have a family of rings ${\cal R}_t={C[x_0,x_1,\ldots, x_n]\over I_t}$ parameterized by $t$.
 The flat limit $I_0 = lim_{t\rightarrow 0} I_t$  is defined as follows. We can decompose any $f \in I$ as $f = f_1+\ldots+ f_k$ into elements in distinct weight spaces for the $C^{*}$ action $\eta$ on $C[x_0, . . . , x_n]$. 
 Let us write $in(f)$ for the element $f_i$ with the smallest weight, which we can think of as the "initial term" of $f$. Then $I_0$ is the ideal generated by the set of initial terms $\{in(f)|f \in I\}$. The test chiral 
 ring is defined as ${\cal R}_0={C[x_0,x_1,\ldots, x_n]\over I_0}$. 
  
 The test chiral ring has the following crucial proerties: a) The flat limit is the same if we use the symmetry generator $s\eta$ with $s>0$; b) $R_0$ is invariant with respect to symmetries of $R$ and $\eta$;
 c): The Hilbert series of ${\cal R}$ and ${\cal R}_0$ are the same for the symmetries of ${\cal R}$, this is the continuity condition on test configuration. 
 Using above procedure, we can get infinite number of  test chiral rings. 
The criteria for determining whether a test chiral ring ${\cal R}_0$ destabilizes a polarized ring $({\cal R},\zeta)$ is  
\begin{definition}
A test chiral ring ${\cal R}_0$ destabilizes $({\cal R},\zeta)$ if ${\cal R}_0$ gives \textbf{no less} central charge $a$ with respect to the space of possible $U(1)_R$ symmetries $a \zeta+s\eta,~s\geq 0$.  
\label{def2}
\end{definition}
It is crucial that $s\geq 0$ so we have the same test chiral ring using the symmetry generator $s\eta$ on ${\cal R}$. 
Now we state the definition of stable chiral ring:
\begin{definition}
A polarized chiral ring $({\cal R},\zeta)$ is called stable if there is no destabilizing test chiral ring. 
\end{definition}
This definition can be thought of as the \textbf{generalized a maximization} procedure. For the original \textbf{a maximization} procedure \cite{Intriligator:2003jj}, we do not change the chiral ring, namely we 
only use the symmetry generator of ${\cal R}$ to generate the test chiral ring, and the flat limit ${\cal R}_0$ is the same as ${\cal R}$.  The hidden assumption in this process is that the ring ${\cal R}$ is already the ring of a SCFT, and we would like to 
determine the correct $U(1)_R$ symmetry. 

Once we define the notion of stability of chiral ring, we would like to state the main conjecture of this paper: 
\begin{conjecture}
A polarized chiral ring $({\cal R},\zeta)$ is the chiral ring of a SCFT if and only if it is stable.
\end{conjecture}
This conjecture answers the question when a chiral ring can be that of a SCFT. 
We would like to test the above conjecture for  general class of $\mathcal{N}=1$ theories.  However, we face several  difficulties. First, it is usually not easy to derive 
the full chiral ring of a theory, and so examples are in short supply.  Secondly, we do not know how to characterize the $U(1)_R$ like symmetry from the chiral ring itself. 
Thirdly, it is not known how to determine the trial central charge $a(\zeta)$ for a symmetry $\zeta$ from the chiral ring itself. 
However, these problems are solved for a class of models arising from string theory.  Precisely, it is possible to determine the  exact chiral ring for $\mathcal{N}=1$ theories derived from 
$N$ D3 branes probing a three dimensional singularity \cite{Klebanov:1998hh}: the 3d singularity can be defined by an affine variety $X$ with coordinate ring ${\cal H}_{X} := \mathbb{C}[x_1,\ldots, x_n]/I$, which
determines the chiral ring ${\cal R}$ of field theory. We can characterize $U(1)_R$ like symmetries by requiring the top form $\Omega$ \footnote{The existence of such form puts restriction on the singularity type.} on $X$ having charge two. 
In the large N limit, the central charge can be computed from the Hilbert series of $X$ \cite{Bergman:2001qi,Martelli:2005tp,Martelli:2006yb}. 

We would like to determine whether the chiral ring $({\cal R},\zeta)$ of above field theory model is stable or not.  Assuming our conjecture relating stability of chiral ring and SCFT, 
the stability of the chiral ring of these models has the following geometric consequence: 
 If the chiral ring is stable, then according to  AdS/CFT dictionary \cite{Maldacena:1997re,Witten:1998qj,Gubser:1998bc}, in the large N limit the IR SCFT is dual to type IIB string 
theory on  $AdS_5\times L_5$ \cite{Klebanov:1998hh, Morrison:1998cs}, where $L_5$ is a five manifold, defined as 
 the link of a 3d singularity $X$, which carries a Sasaki-Einstein (SE) metric. The $U(1)_R$ symmetry $\zeta$ is identified with the Reeb vector field on $L_5$. 
 In other words, the stability of the chiral ring is equivalent to the existence of Sasaki-Einstein metric on $L_5$, or equivalently 
 the existence of a Ricci-flat conic metric on $X$.

The existence of Sasaki-Einstein metrics has been studied extensively recently in mathematics literature.  Briefly, in the setting of Fano K\"ahler manifolds, the Yau-Tian-Donaldson conjecture predicted that the existence of  K\"ahler-Einstein metrics with positive scalar curvature is equivalent to the algebro-geometric notion of K-stability \cite{donaldson2002scalar} which is an improvement of the original conjecture 
of Yau \cite{yau1993open}.  
This conjecture was recently proved by Chen-Donaldson-Sun \cite{chen2015kahler,chen2015kahler1,chen2015kahler2}.  In our more general context, a notion of K-stability and it's implications for the existence 
of Sasaki-Einstein metrics was studied by the first author and Sz\'ekelyhidi in \cite{collins2012k,collins2015sasaki}.  The notion of K-stability involves constructions of so called test configurations ${\cal X}$ and the criteria for determining whether ${\cal X}$ destabilizes $X$ is determined by the sign of the so-called Donaldson-Futaki invariant.
One of major point of this paper is to provide an interpretation of the  Donaldson-Futaki invariant as a version of generalized $a$ maximization: 
\begin{theorem}
The K-stability of the affine variety $X$ is equivalent to the stability of the chiral ring of the corresponding field theory. 
\end{theorem}

The paper is organized as follows: section two reviews some basic facts about $\mathcal{N}=1$ chiral ring; section three studies theory engineered using
D3 brane probing certain three dimensional singularity $X$, and  K-stability of $X$ is interpreted as the generalized $a$ maximization procedure introduced above; section four discusses
some physical consequences from K stability; finally, a conclusion is given in section five.

\section{Generality of chiral ring}
Consider a four dimensional $\mathcal{N}=1$ supersymmetric field theories. A  chiral operator ${\cal O}_i$ is defined as an operator annihilated by supercharges $\bar{Q}_{\dot{\alpha}}$, and 
is defined modulo cohomology of $\bar{Q}_{\dot{\alpha}}$: ${\cal O}_i\sim {\cal O}_i+[\bar{Q}_{\dot{\alpha}}, \chi]$ \cite{Cachazo:2002ry}. Chiral operators  have some interesting properties:
\begin{itemize}
\item The sum of two chiral operators is still a chiral operator, and the product of two chiral operators is still a chiral operator. 
\item There is an identity operator.
\item The expectation value of a product of chiral operators are independent of their positions, and they have the simple OPE structure ${\cal O}_i{\cal O}_j=\sum C_{ij}^k {\cal O}_k$ with $C_{ij}^k$ constant. 
\end{itemize}
These properties imply that the chiral operators form a commutative ring with an identity. To solve a $\mathcal{N}=1$ theory,
one would like to determine the full set of chiral operators.  That is, one would like to find the generators and relations determining the chiral ring. For example, for $SU(N)$ gauge theory with an adjoint matter field $\Phi$, the generators of single trace chiral operators are: $\text{Tr}(\Phi^k)$, $\text{Tr}(W_\alpha \Phi^k)$, and $\text{Tr}(W_\alpha W^\alpha\Phi^k)$, 
the full chiral ring relations are determined in \cite{Cachazo:2002ry}. The chiral ring relations usually are much harder to determine. Typically, we have the following classical chiral ring relations;
\begin{itemize}
\item  Chiral ring relations  
come from the finite size of matrices, and we have Caley-Hamilton equation for a matrix. For example, 
for a chiral field in the adjoint representation of gauge group $SU(N)$, the chiral operators $\text{Tr}(\phi^i), i> N$ can be expressed in terms of  $\text{Tr}(\phi^j)$ 
 with $j\leq N$. 
 \item  Chiral ring relations come  from the constraints of the superpotential. For example, consider
$\mathcal{N}=4$ $SU(N)$ gauge theory: this theory has three chiral fields $X, Y, Z$, and a superpotential $W=\text{Tr}XYZ-\text{Tr} X Z Y$. The $F$-term equations from the superpotential are
 \begin{equation}
 [X,Y]=[Y,Z]=[Z,X]=0,
 \end{equation}
and so the matrices $X,Y,Z$ commute, which lead to chiral ring relations of the type $\text{Tr}(XYZ)=\text{Tr}(XZY)$ and so on. 
 \end{itemize}
These classical chiral ring relations can be modified by quantum effects such as instantons, Konishi anomalies and strongly coupled dynamics, and we have the quantum chiral ring.
The determination of the generators of chiral ring and the quantum chiral ring relation is a central task in the study of supersymmetric gauge theory. 
Let's assume that the generators of the chiral ring are $x_1, x_2, \ldots x_s$, and the chiral ring relations are generated by polynomial relations, 
then the quantum chiral ring is isomorphic to
\begin{equation}
\mathbb{C}[x_1, x_2,\ldots, x_s]/I,
\end{equation}
where $\mathbb{C}[x_1, x_2,\ldots, x_s]$ is ring of polynomials with complex coefficients, and $I$ is the ideal generated by the chiral ring relations.   
In general,  the parameters of our theory such as the dynamically 
generated scale $\Lambda$ and masses $m_i$ should be included into the generators and relations of chiral ring. From now on, by the chiral ring we will always mean the quantum chiral ring.

\textbf{Example}: Consider $\mathcal{N}=1$ $SU(N)$ SQCD with $N_f=N$ quarks $Q_i, \tilde{Q}_j$, the space of
chiral operators are
\begin{align}
& M_{ij}=Q_i\tilde{Q}_j,  \nonumber\\
&B=\epsilon_{\alpha_1 \alpha_2 \ldots \alpha_N} Q_1^{\alpha_1}\ldots Q_N^{\alpha_N},  \nonumber\\
&\tilde{B}=\epsilon_{\alpha_1 \alpha_2 \ldots \alpha_N} \tilde{Q}_1^{\alpha_1}\ldots \tilde{Q}_N^{\alpha_N}.
\end{align}
The classical chiral ring relation is $\text{Det}(M)-B\tilde{B}=0$, but quantum mechanically, the ring relation is changed to 
\begin{equation}
f=\text{Det}(M)-B\tilde{B}-\Lambda^{2N_c}=0.
\end{equation}
Here $\Lambda$ is the dynamical scale of the theory.  The chiral ring is then $C[M, B, \tilde{B},\Lambda]/f$ \cite{Intriligator:1995au}.

We are interested in the chiral ring of a SCFT. The $\mathcal{N}=1$ SCFT has a distinguished $U(1)_R$ symmetry, and the scaling dimension of a  chiral operator are related to
its $U(1)_R$ charge by
\begin{equation}
D({\cal O})={3\over 2} R({\cal O}).
\end{equation} 
In general, it is not easy to determine the $U(1)_R$ symmetry of a SCFT. Intriligator and Wecht found a remarkable $a$-maximization procedure to determine the $R$-symmetry \cite{Intriligator:2003jj}.  Namely, they predicted that the correct $R$-symmetry maximizes the central charge $a$. 

One usually defines a SCFT as the IR limit of a UV quiver gauge theory, and the $U(1)_R$ symmetry of IR SCFT can be determined as follows: 
First, find all the anomaly free $U(1)$ symmetries in the UV and 
define a trial R symmetry $U(1)_{trial}=\sum_I s_I F_I$.   Second, compute the trial central charge using the formula
\begin{equation}
a_{trial}(s_I)={3\over 32} (3 \text{Tr}(R_{trial}^3)-\text{Tr}(R_{trial})).
\end{equation}
The true $U(1)_R$ symmetry is found by maximizing the central charge and this will fix the coefficients $s_I$. 
A crucial assumption of above procedure is that all the symmetries for the IR SCFT are manifest in the UV description! 
However, this is often not the case.  For example, two possible scenarios are
 \begin{itemize}
 \item The violation of unitarity bound: if  a gauge invariant operator ${\cal O}$ violates the unitarity bound after doing $a$-maximization,  it is 
 argued that this field becomes free \cite{Seiberg:1994bz}, and that there is an accidental $U(1)$ symmetry acting on this operator ${\cal O}$ only.
 \item Even if there is no violation of unitarity bound, accidental symmetry is still possible as a result of some unknown dynamical effect. 
  \end{itemize}
 
 \textbf{Example}: Let's illustrate the above point by an example. Consider $\mathcal{N}=1$ $SU(N_c)$ SQCD with $N_f$ fundamental flavors. There is an unique $U(1)_R$ type symmetry
 such that the quarks $Q$ and antiquarks $\tilde{Q}$ have the following charges
 \begin{equation}
 R_{Q}=R_{\tilde{Q}}={N_f-N_c\over N_f}.
 \end{equation}
 We require $N_f>N_c$ so that the $R$ charge is positive. The mesons  has $R$ charge $R(M)=2{N_f-N_c\over N_f}$, and baryons an anti-baryons have 
 $R$ charge $R(B)=R(\tilde{B})={N_c(N_f-N_c)\over N_f}$. So using this candidate $U(1)_R$ symmetry for the IR SCFT, we have $\Delta(M)={3\over2}R(M)<1$ 
 if $N_c<N_f<{3\over 2} N_c$ and it is argued that these mesons become free in the IR \cite{Seiberg:1994bz}.
 The baryons do not violate the unitarity bound, however,  it becomes free if $N_f=N_c+1$ as we can see it from Seiberg dual 
 description \cite{Seiberg:1994pq}. 
 
 In fact, the appearance of accidental symmetry implies that the chiral ring of the IR SCFT is different from the UV theory, which means that the chiral ring of UV theory is not stable. 
 Since it is difficult to detect the appearance of accidental symmetry, it is also difficult to tell whether the UV chiral ring is stable or not. 
 In the next section, we will consider a class of $\mathcal{N}=1$ models where we will relate the stability of chiral ring to a problem in geometry.

 \section{K-stability and stability of chiral ring}
 
 \subsection{The chiral ring, Hilbert series and the central charge $a$ }
\begin{figure}[h]
\centering
  \includegraphics{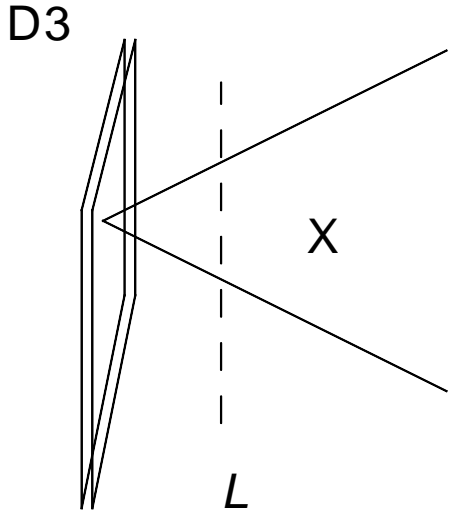}
  \caption{One can engineer four dimensional $\mathcal{N}=1$ theories using D3 brane probing three dimensional klt  Gorenstein singularity $X$. For the singularity $X$, one can define a 
  link $L$ which is a five dimensional Sasakian manifold.}
  \label{d2}
  \end{figure}

Consider a $\mathcal{N}=1$ theory on world volume of  $N$ D3 branes probing a graded three dimensional normal, Kawamata log-terminal (klt), Gorenstein singularity $X$, (see figure. \ref{d2}).  The 3d singularity is defined by an affine ring
\begin{equation}
{\cal H}_X=\mathbb{C}[x_1, x_2, \ldots, x_r]/I,
\end{equation}
here $\mathbb{C}[x_1, x_2, \ldots, x_r]$ is the polynomial ring and $I$ is an ideal. Let's explain the meaning for various terms characterizing our singularity: \textbf{normal} means that the 
codimension of the singular locus $P$ of $X$ is no less than two; \textbf{graded} means that there is at least one $C^*$ action on $X$; \textbf{Gorenstein} means that the canonical sheaf $K_X$ is 
a line bundle and one has a non-vanishing top form $\Omega$ on $X/P$;  \textbf{Kawamata log-terminal} can be characterized that the volume form $\Omega\wedge\overline{\Omega}$ has finite mass near the singularities of $X$ (see \cite{collins2015sasaki}).

We have the following map between the properties of the ring $X$ and field theory,
\begin{itemize}
\item The automorphism group $G$ of $X$  gives the (complexified) anomaly free symmetries of the field theory, and the possible $U(1)_R$ symmetry $\zeta$ is a subgroup of $G$.
\item The coordinate ring of the vacua moduli space is described as the coordinate ring of the variety $M_{N}=X^N/S^N$. In the large N limit, the single trace operators parameterizing $M_\infty$ can be identified as  the ring elements of $X$ \footnote{Notice that in the large N limit the ring structure of $X$ is not 
the chiral ring structure of the field theory. In the large N limit, the chiral ring structure is trivial, namely the product of two single trace operator defines the multiple trace operators.}: namely the holomorphic 
functions on $X$ give the chiral scalar operators of the field theory in the large N limit. So $X$
essentially determines the nontrivial part of the chiral ring \footnote{There are other types of scalar chiral operators  which do not get expectation value, and also chiral Baryonic operators. These operators seem not affect the stability issue of our model.}. 
\item $X$ has a canonical $(3,0)$ form $\Omega$, and it has charge 2 under the \textbf{possible} $U(1)_R$ symmetry $\zeta$:
\begin{equation}
[\Omega]=2,
\end{equation}
(in fact, this condition is equivalent $X$ being klt). We also require that the $U(1)_R$ charge of the coordinates $x_i$ is positive. 
\end{itemize}

Consider a possible $U(1)_R$ symmetry $\zeta$ which is realized as an automorphism of $X$.  The trial central charge $a(\zeta)$ (of order $N^2$) of the field theory can be computed from the Hilbert series of the ring $X$ \cite{Bergman:2001qi,Martelli:2005tp,Martelli:2006yb,Eager:2010yu}. The Hilbert series of $X$ with respect to $\zeta$ is defined by 
\begin{equation}
Hilb(X,\zeta,t)=\sum (dim H_{\alpha}) t^\alpha;
\end{equation}
Here $H_\alpha$ is the subspace of ring ${\cal H}_X$ with charge $\alpha$ under the action $\zeta$.  The Hilbert series has a Laurent series expansion around $t=1$ obtained by setting $t=e^{-s}$ and expanding
\begin{equation}
Hilb(X,\zeta, e^{-s})={a_0(\zeta)\over s^3}+{a_1(\zeta)\over s^2}+\ldots
\end{equation}
The coefficients $(a_0(\zeta), a_1(\zeta))$ have  following properties:
\begin{itemize}
\item  $a_0$ is proportional to the volume of the link $L_5$ of the singularity, and the trial central charge $a(\zeta)$ (order $N^2$ term) is related to $a_0$ as 
\begin{equation}
a(\zeta)={27 N^2\over 32} {1\over a_0(\zeta)}.
\label{cent}
\end{equation}
\item $a_0=a_1$ which is due to the condition that $\Omega$ has charge $2$.
\item $a_0$ is convex function of the symmetry generators \cite{Martelli:2006yb}. 
%\item If the ring $X$ is stable, the true $U(1)_R$ symmetry of the field theory is found by minimizing $a_0$ (maximizing the central charge $a$) . 
\end{itemize}

For the singularity $X$, one can define a 5 dimensional link $L_5$ with Sasakian structure \cite{boyer2008sasakian}. 
If there is a Sasaki-Einstein metric on the link $L_5$, one can find the true $U(1)_R$ symmetry by minimizing $a_0$, and the field theory 
central charge is given by the formula~\eqref{cent}. In the large N limit, the SCFT on D3 branes is dual to Type IIB string theory on the following geometry 
\begin{equation}
AdS_5\times L_5. 
\end{equation}
The existence of the SE metric on $L_5$ is also equivalent to the existence of a Ricci-flat conic metric on $X$.

\textbf{Example}: Consider the conifold singularity defined by the principal ideal $f(z)=z_0^2+z_1^2+z_2^2+z_3^2=0$, and it is known that the link $L_5$ is the manifold $T^{1,1}$ and has a Sasaki-Einstein metric. 
There is 
a $C^*$ action $\zeta$ on this singularity $f(\lambda^{q_i} z_i)=\lambda f(z_i)$ with weights $({1\over 2},{1\over 2},{1\over 2},{1\over 2})$. 
The canonical three form is $\Omega={dz_0\wedge d z_1\wedge d z_2\wedge dz_3\over dF}$. $\Omega$ has charge $1$ under the symmetry $\zeta$, so the 
possible $U(1)_R$ symmetry is actually $\zeta^{'}=2\zeta$ in order to ensure $\Omega$ has charge two.
The Hilbert series of $X$ with respect to symmetry generator $\zeta^{'}$ is
\begin{equation}
Hilb(t)={(1-t^2)\over (1-t)^4}|_{t=e^{-s}}={2\over s^3}+{2\over s^2}+\ldots
\end{equation}
Using formula \ref{cent}, We find that the central charge is equal to $a={27\over 64} N^2$ which agrees with the result derived from field theory \cite{Klebanov:1998hh}.

\subsection{K Stability and generalized $a$-maximization}
Now thel question is whether the link $L_5$ has Sasaki-Einstein metric. This question is reduced to studying the K-stability of the ring $X$ \cite{collins2012k,collins2015sasaki}. 
On the other hand, $X$ essentially determines the chiral ring of the field theory, and if the chiral ring of the field theory is stable, i.e. it is a chiral ring of 
a SCFT, the field theory is dual to type IIB string theory on the background $AdS_5\times L_5$, where $L_5$ has a Sasaki-Einstein metric. From this AdS/CFT correspondence, 
one can see that K-stability should be equivalent to the stability of the chiral ring of field theory defined in introduction. 
In this subsection, we will discuss two crucial ingredients of K-stability;  test configuration and the Donaldson-Futaki invariant. We will also give a physical interpretation of these two elements and 
show that K-stability is equivalent to the stability of the chiral ring. 

\subsubsection{Test configurations}
\begin{figure}[h]
\centering
  \includegraphics{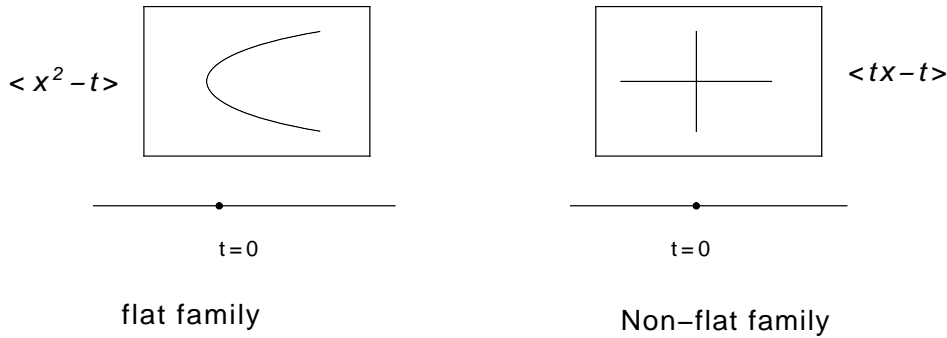}
  \caption{Left: A flat family of rings. At $t\neq 0$, there are two points and the configuration degenerates into one point at $t=0$, which is the central fibre of this flat limit. Right: A non-flat family of rings. At $t\neq 0$, the ring is zero dimensional, but at $t=0$, the ring is one dimensional. }
  \label{flat}
  \end{figure}

Let's first describe the definition of a test configuration arising in K stability, which  actually motivates our definition of test chiral ring in the introduction.  
In the K-stability context, one constructs a test configuration by constructing 
a flat family $\pi: {\cal X} \rightarrow \mathbb{C}$ (for a simple illustration of flat and non-flat family, see figure. \ref{flat}.). This flat family is generated by a one dimensional symmetry generator $\eta$, and for $t\neq 0$, the ring ${\cal}H_{X_t}$ corresponding to the fiber $X_t = \pi^{-1}(t)$ is isomorphic to the original ring ${\cal H}_{X}$. At $t=0$, the ring degenerates into 
a different ring which we call $H_{X_0}$, and it is also called central fibre. 

The flat limit is a quite common concept in algebraic geometry, but its definition is quite involved and we do not want to give a detailed introduction here.  For the interested reader, see section 6 of 
\cite{eisenbud2013commutative}. Here, we just want to  point out  several important features of the flat family constructed above.
\begin{itemize}
\item[(a)] The Hilbert series is not changed if we use the same symmetry generator for the new ring ${\cal H}_{X_0}$.  In particular,  
$X_0$ has the same dimension as $X$.
\item[(b)] The maximal torus in the automorphism group of the central fibre $X_0$ has one more dimensional symmetry generated by $\eta$, unless $X_0 \cong X$. 
\end{itemize}
We require that the degeneration is normal (which implies that the codimension of the singular locus is not less than two). The new singularity $X_0$ is still Gorenstein and klt and, in the non-trivial case, possesses an extra one-dimensional symmetry. 

\textbf{Example}: Consider the ring $X$ defined by the ideal $x^2+y^2+z^2+w^k=0$, and consider a $\mathbb{C}^*$ action $\eta$ which acts only on coordinate $w$ with the action $\eta(w)=t w$. We then get a family of 
rings parametrized by the coordinate $t$:
\begin{equation}
x^2+y^2+z^2+t^kw^k=0.
\end{equation} 
The flat limit of this family over $t=0$ is found (in this case) by keeping the terms with lowest order. The central fiber of this test configuration is then cut out by the equation
\begin{equation}
x^2+y^2+z^2=0
\end{equation}
Notice that $l\eta$ with $l>0$ gives the same degeneration limit $X_0$. On the other hand $l\eta$ with $l<0$ gives a different degeneration limit-- we get the ring generated by the ideal $w^k=0$, which is not normal!

\subsubsection{Futaki invariant and generalized $a$-maximization}
Now let's start with a ring $X$ with symmetry $\zeta$ and we also choose the generators  $t_i, i=1,\ldots, n$ for the Lie algebra $\mathfrak{t}$ of the maximal torus in the automorphism group $G$ of X.  Let us write 
$\zeta=\sum_{i=1}^n \zeta_it_i$, and we may as well assume that $\zeta$ minimizes the volume over all the possible $U(1)_R$ symmetries parametrized by $\mathfrak{t}$. 
Consider a test configuration ${\cal X}$ generated by a symmetry generator $\eta$ and let $X_0$ denote the central fibre. We would like to determine whether or not $X_0$ destabilizes $X$. The 
crucial ingredient is the Donaldson-Futaki invariant defined in \cite{donaldson2002scalar}. 

The ring $(X_0, \zeta, \eta)$ is still Gorenstein and klt, and has a at least two dimensional symmetry group generated by $\zeta$ and $\eta$. There is only  
one dimensional possible $U(1)_R$ symmetry as we need to impose following two conditions
\begin{itemize}
\item[(a)] The charge on the coordinates $x_i$ is positive
\item[(b)] The  $(3,0)$ form has charge 2.  
\end{itemize}
The second condition can be fixed by computing the Hilbert series of $X_0$ with respect to  symmetry generator and  
imposing the condition $a_0=a_1$. This one dimensional candidate $U(1)_R$ symmetry can be parameterized as
\begin{equation}
\zeta(\epsilon)=\zeta+\epsilon(\eta-a\zeta).
\end{equation}
Notice that we require $\epsilon>0$ so that the central fibre is the same as the original one if we use the symmetry $\epsilon(\eta-a\zeta)$ to generate the test configuration. 
Substitute the above parameterization into the equation $a_0=a_1$ and expand it to first order in $\epsilon$,  we have
\begin{equation}
a_0(\zeta+\epsilon(\eta-a\zeta))=a_1(\zeta+\epsilon(\eta-a\zeta)) \rightarrow a_0(\zeta)+\epsilon (\eta-a \zeta)\cdot a_0^{'} = a_1(\zeta)+\epsilon (\eta-a \zeta)\cdot a_1^{'}.
\end{equation}
Here $a_0^{'}$ and $a_1^{'}$ are the vectors defined by the derivative ${da_i(\vec{x})\over d {\vec{x}}}|_{\vec{x}=\zeta}$, and $\vec{x}=\sum_{i=1}^ns_i t_i+b \eta$. Using the result $a_0(\zeta)=a_1(\zeta)$, we have 
\begin{equation}
a={\eta\cdot (a_0^{'}-a_1^{'}) \over \zeta \cdot (a_0^{'}-a_1^{'})}={\eta\cdot (a_1^{'}-a_0^{'}) \over a_0}={1\over a_0(\zeta)}({da_1(\zeta+\epsilon \eta) \over d \epsilon }-{d a_0(\zeta+\epsilon \eta) \over d\epsilon })|_{\epsilon=0}.
\label{constant}
\end{equation} 
We also use the fact $\zeta\cdot a_0^{'}=3 a_0(\zeta),~\zeta\cdot a_1^{'}=2a_1(\zeta)=2a_0(\zeta)$. 
Now the Futaki invariant is defined to be 
\begin{equation}
F(X,\zeta, \eta)=D_\epsilon a_0(\zeta(\epsilon))|_{\epsilon=0}.
\end{equation}
This definition is not of the form of the original Futaki invariant defined in \cite{donaldson2002scalar}, however, we will now show that our definition is equivalent to the original one (see also \cite{collins2015sasaki} for more discussion). We have
\begin{align}
& F(X,\zeta,\eta)=D_\epsilon a_0(\zeta+\epsilon(\eta-a\zeta))=(\eta-a \zeta) \cdot a_0^{'} \nonumber\\
&=D_\epsilon a_0(\zeta+\epsilon \eta)-{\eta\cdot (a_0^{'}-a_1^{'}) \over a_0}D_\epsilon a_0(\zeta+\epsilon \zeta) \nonumber\\
&=D_\epsilon a_0(\zeta+\epsilon \eta)+3a_0(\zeta){\eta\cdot (a_0^{'}-a_1^{'}) \over a_0} \nonumber \\
&= D_\epsilon a_0(\zeta+\epsilon \eta)+3a_0(\zeta)D_{\epsilon}{a_1(\zeta+\epsilon \eta)\over a_0(\zeta+\epsilon \eta)}. 
\label{futaki}
\end{align}
We use the definition from first line to second line, and from second line to third line we use the fact $D_\epsilon a_0(\zeta+\epsilon \zeta)=-3 a_0(\zeta)$ (which can be found using the definition of Hilbert series). 
The formula in the last line is precisely the Futaki invariant defined in \cite{collins2015sasaki}.  Having defined the Futaki invariant, we can now state the definition of K-stability.

\begin{theorem}
A polarized ring $(X,\zeta)$ is stable if for any non-trivial test configuration generated by the symmetry $\eta$, the Futaki invariant satisfies
\begin{equation}
F(X, \zeta, \eta)>0.
\end{equation}
And for the trivial test configuration, namely the central fibre $X_0$ is the same as $X$, the Futaki invariant satisfies
\begin{equation}
F(X, \zeta, \eta)\geq0.
\end{equation}
\end{theorem}
We now provide a physical interpretation of the Futaki invariant $F$. Since $a_0$ is inverse proportional to the central charge of the coordinate ring of the central fiber, the Futaki invariant is directly related to the maximization of the central charge. The shape of the function $a_0$ with respect to $\epsilon$ is 
drawn in figure. \ref{futaki}. $F<0$ implies that  $a_0(\epsilon)$ is minimized at $\epsilon>0$, and the new ring gives larger central charge $a$!
When $F=0$ and $X_0$ is different from X, the two ring gives the same central charge $a$, but the central fiber $X_0$ has a strictly larger symmetry group which then 
destabilizes $X$. When $F>0$, the new ring gives less central charge over the allowed space of symmetries ($\epsilon>0$). In summation, Futaki invariant is actually implying generalized $a$-maximization.  Namely, a test configuration $X_0$ destabilizes $X$ if it gives no less central charge!
\begin{figure}[h]
\centering
  \includegraphics{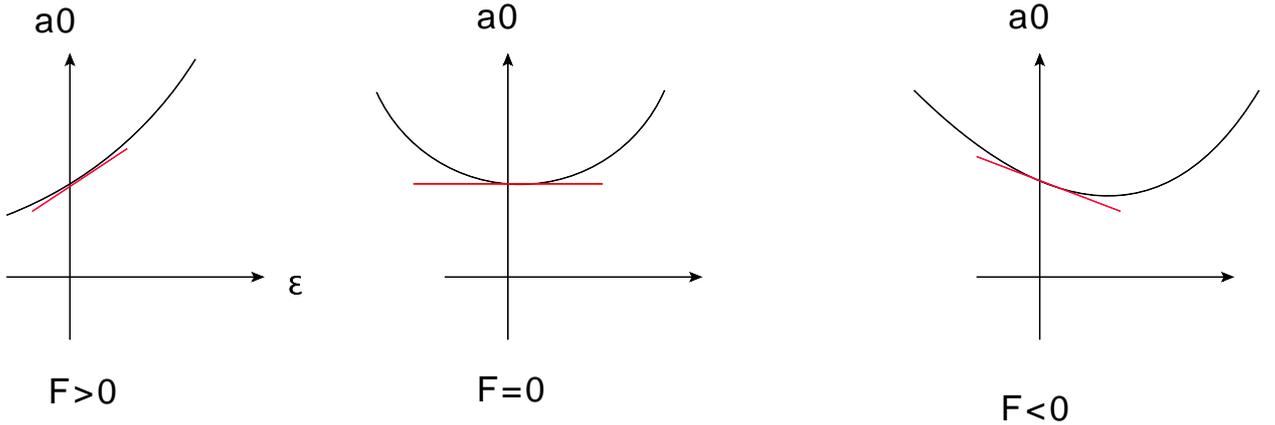}
  \caption{Three situations for Futaki invariant. $F>0$: $a_0(\epsilon)>a(0)$ for $\epsilon>0$; $F=0$: the minima of $a_0$ is achieved at $\epsilon=0$; $F<0$: the minima of $a_0$ is  achieved fro $\epsilon>0$. 
   Notice that we only need to look at $a_0$ for $\epsilon>0$.}
  \label{futaki}
  \end{figure}

\textbf{Example}: Consider the ring $X$ which is generated by the ideal $x^2+y^2+z^2+w^k=0$, this ring has a symmetry $\zeta$ with charge $({2k\over k+2},{2k\over k+2},{2k\over k+2},{4\over k+2})$ on coordinates $(x,y,z,w)$. This symmetry is chosen such that the 
$(3,0)$ form $\Omega={dx \wedge dy \wedge dz \wedge dw\over df}$ has charge two. The Hilbert series for $\zeta$ is 
\begin{equation}
Hilb(X,\zeta, t)=\frac{1-t^{\frac{4 k}{k+2}}}{\left(1-t^{\frac{4}{k+2}}\right) \left(1-t^{\frac{2 k}{k+2}}\right)^3}.
\end{equation}
Expand around $t=1$, we find $a_0(\zeta)=a_1(\zeta)={(2+k)^3\over 8k^2}$. 
Now consider the test configuration generated by the symmetry $\eta$ with charges $(0,0,0,1)$.  In this case, the central fibre $X_0$ is generated by the
ideal $x^2+y^2+z^2=0$. Using formula \eqref{constant}, 
the one parameter possible $U(1)_R$ symmetry is 
\begin{equation}
\zeta(\epsilon)=\zeta+\epsilon(\eta-{1\over 2}\zeta).
\end{equation}
The Hilbert series with respect to above symmetry is 
\begin{equation}
Hilb(X_0,\zeta(\epsilon),t)={1-t^{(1-{\epsilon\over 2}){4k\over k+2}}\over (1-t^{(1-{\epsilon\over 2}){2k\over k+2}})^3(1-t^{(1-{\epsilon\over 2}){4\over k+2}+\epsilon)})}.
\end{equation}
Substituting $t=exp(-s)$ and expand the Hilbert series around $s=0$, we get 
\begin{equation}
a_0(\zeta(\epsilon))=a_1(\zeta(\epsilon))=\frac{2 (k+2)^3}{(\epsilon-2)^2 k^2 (\epsilon k+4)}.
\end{equation}
The Futaki invariant is computed as 
\begin{equation}
F=D_\epsilon a_0(\zeta(\epsilon))|_{\epsilon=0}=\frac{(4-k) (k+2)^3}{32 k^2}.
\end{equation}
So $F\leq 0$ for $k\geq 4$.  Since $X_0$ is clearly not isomorphic to $X$, we conclude that $X_0$ destabilizes $X$ for $k\geq 4$. A physical interpretation of this result will be given in the next section. 

\subsubsection{Some discussions}
Checking K-stability involves two steps.  First, finding a test configuration and then computing the Futaki invariant.  While the computation of Futaki invariant is straightforward, the set of possible test configurations is in principle infinite.  Thus, in order to check K-stability one needs to reducing the 
sets of possible test configurations. There are several simplifications we can make
\begin{itemize}
\item The first simplification has already been used, namely we require that the central fibre to be normal, Gorenstein and klt. This is simply due to the reason that the central fibre should describe the chiral ring of 
a $\mathcal{N}=1$ field theory. 
\item Assume that the symmetry group of the ring $X$ is $G$, then one only need to consider the flat families generated by a symmetry which commutes with $G$ \cite{datar2015k, collins2015sasaki}. This fact is quite useful 
for singularities with many symmetries. In particular, if the variety has three dimensional symmetries (or in other words, $X$ is toric), then there are no non-trivial test configurations, and hence checking stability reduces to volume minimization (or $a$-maximization). 
\end{itemize}

\section{Some physical consequences}

\subsection{$a$-maximization}
Let's assume that the ring $X$ is stable and has more than one dimension worth of possible $U(1)_R$ symmetry. The determination of $U(1)_R$ symmetry is 
solved by $a$-maximization \cite{Intriligator:2003jj} or equivalently volume minimization \cite{Martelli:2005tp,Martelli:2006yb}. We now show that $a$-maximization can be explained using K-stability. 
Consider a test configuration generated by the symmetry vector $\eta$, and the central fibre $X_0$ is the same as $X$.
The Futaki invariant is 
\begin{equation}
F=D_\zeta a_0(\zeta+\epsilon \eta)|_{\epsilon=0}=\eta\cdot a_0^{'}(\zeta),
\end{equation}
If $F(X, \zeta, \eta)>0$, the test configuration $(X_0, \zeta, \eta)$ does not destabilize $X$.  But, since $\eta$ preserves $X$,  we can use the symmetry generator $-\eta$ to generate a test configuration with the same central fibre $X$, 
and the Futaki invariant now is  $F(X, \zeta, -\eta)=-F(X, \zeta, \eta)<0$ which will make the ring unstable. So K-stability implies that the symmetry generator has to satisfy 
\begin{equation}
a_0^{'}(\zeta)=0.
\end{equation}
Notice that since $a_0$ is a convex function, the solution of above equation is the minimum, and therefore the central charge $a$ is maximized.

\subsection{Unitarity bound}
One can always generate a test chiral ring by using a symmetry acting on a single coordinate $x$ only. The central 
fibre $X_0$ is a new ring with $x$ free \footnote{Mathematically such test configuration is generated by the so-called Rees algebra.}. The 
Futaki invariant is computed in \cite{collins2012k}, and the answer is 
\begin{equation}
F\propto (dim(x)-1),
\end{equation}
here we ignore a positive constant, and $dim(x)$ is the scaling dimension of the chiral scalar operator $x$. $X$ is not 
destabilized by this particular test configuration if 
\begin{equation}
dim(x)>1.
\end{equation}
This is nothing but the unitarity bound on scalar operator represented by $x$. 

\subsection{Singularity with more than one dimensional symmetries}
Consider a toric Gorenstein singularity, and the rank of symmetry group is $3$. As we discussed above, there is no non-trivial 
test configuration, and so toric singularity is stable provided we choose the $U(1)_{R}$ symmetry which minimizes the volume. 
On the other hand, the existence of Sasaki-Einstein metrics on the link of a toric singularity was established using analytic methods in \cite{futaki2009transverse}.
For the ring $X$ with two dimensional symmetries, one only needs to check finite number of test configurations, see \cite{collins2015sasaki}.

\textbf{Example}: Consider the ring defined by the ideal $x^2+y^2+z^p+w^q=0$. This ring has a two dimensional 
symmetry group. One can characterize all the test chiral rings, and the ring is proven to be stable if $(p,q)$ satisfies the following condition \cite{collins2015sasaki}: 
\begin{equation}
p<2q~\text{and}~q<2p.
\end{equation} 
Notice that this is just the requirement of the unitarity bound on the operators represented by $z$ and $w$.  However, as we will see later, the unitarity bound is not the only
obstruction which can appear, even in the case of hypersurface singularities.

\subsection{Hypersurface singularity}
Consider an isolated three-fold hypersurface singularity $f:(C^4,0)\rightarrow (C,0)$ with a $C^*$ action $\vec{\zeta}$:
\begin{equation}
f(\lambda^{w_i}q_i)=\lambda f(z_i). 
\end{equation}
Here all the charges $w_i$  are positive. The canonical three-form is 
\begin{equation}
\Omega={dz_0\wedge dz_1\wedge dz_2\wedge dz_3\over d F}.
\end{equation}
This form has charge $\sum w_i-1$, and the candidate $U(1)_R$ symmetry is found by requiring $\Omega$ to have charge two:
\begin{equation}
(\sum w_i-1)\delta=2\rightarrow \delta={2\over \sum w_i-1},
\end{equation}
and the candidate $U(1)_R$ symmetry is $\zeta^{'}=\delta \zeta$. 
To make the coordinate $z_i$ have positive r charge, we require $\sum w_i-1>0$, which implies that the singularity is a rational Gorenstein (and hence klt) singularity. 
Such rational hypersurface singularities have been classified by Yau and Yu \cite{yau2003classification}. 

The Hilbert series of a hypersurface singularity is easy to compute. It takes the following form
\begin{equation}
Hilb(f,t,\zeta^{'})={1-t^{\delta}\over (1-t^{w_0 \delta})(1-t^{w_1 \delta})(1-t^{w_2 \delta})(1-t^{w_3 \delta})}.
\end{equation}

Now let's consider a test configuration which is derived by using a one parameter transformation $\eta$. For simplicity, let's assume that 
the action $\vec{\eta}$ is diagonal on the coordinates with charges $(v_1,v_2,v_3,v_4)$.  In the flat limit, we get a new polynomial $f_0$ which does not 
necessarily define an isolated singularity.  $f_0$ has two dimensional symmetries generated by $(\zeta^{'}, \eta)$, however, there is only a one dimensional 
symmetries which could be the possible $U(1)_R$ symmetry. The one parameter symmetry group can be parameterized as
\begin{equation}
\eta(\epsilon)=\zeta^{'}+\epsilon(\eta-a \zeta^{'}),
\end{equation}
and $a$ can be computed using the formula \ref{constant}.
The Futaki invariant can be computed using formula \ref{futaki}, and we have
\begin{align}
&F(f,\zeta^{'}, \eta)=D_{\epsilon} a_0(\zeta(\epsilon))|_{\epsilon=0}= \nonumber\\
&-[(v_4w_1w_2w_3(w_1+w_2+w_3-2w_4-1)+v_3 w_1 w_2 w_4(w_1+w_2+w_4-2w_3-1)+ \nonumber\\
&v_2w_1w_3w_4(w_1+w_3+w_4-2w_2-1)+v_1w_2w_3w_4(w_2+w_3+w_4-2w_1-1).
\label{hyper}
\end{align}
In the following, we are going to use this formula to test whether a hypersurface singularity is stable or not. 

\subsubsection{Irrelevance of superpotential term}
Recall that we have already studied the singularity 
\begin{equation}
f=z_0^2+z_1^2+z_2^2+z_3^{2k},
\end{equation}
from K-stability perspective, and we  showed  that this ring is unstable for $k\geq 2$. The destabilizing configuration has the central fibre $X_0= \{z_0^2+z_1^2+z_2^2=0\}$; see section $3.2.2$. 

Let's interpret this result from field theory point of view. The quiver gauge theory description is found in \cite{Cachazo:2001sg}; see 
figure \ref{quiver} below. 
\begin{figure}[h]
\centering
  \includegraphics{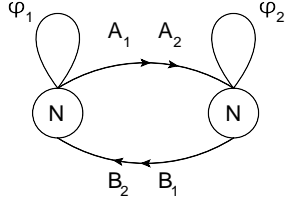}
  \caption{Quiver gauge theory description for D3 brane probing the singularity defined by $z_0^2+z_1^2+z_2^2+z_3^{2k}=0$.  The superpotential is described in \eqref{super}.}
  \label{quiver}
  \end{figure}
We have the following superpotential term:
\begin{equation}
W=\text{Tr} (\phi_1(A_1B_1+A_2B_2))-\text{Tr}(\phi_2 (B_1A_1+B_2 A_2))-2{\text{Tr}\phi_1^{k+1}\over k+1}+2{\text{Tr}\phi_2^{k+1}\over k+1}.
\label{super}
\end{equation}
The $U(1)_R$ charge is fixed such that the NSVZ $\beta$ function is zero, and each term in superpotential $W$ has charge two:
\begin{align}
&{1\over2}[(R(A_1)-1)+(R(A_2)-1)+(R(B_1)-1)+(R(B_2)-1)]+(R(\phi_1)-1)+1=0, \nonumber\\
&{1\over2}[(R(A_1)-1)+(R(A_2)-1)+(R(B_1)-1)+(R(B_2)-1)]+(R(\phi_2)-1)+1=0, \nonumber\\
&R(A_1)+R(B_1)+R(\phi_1)=2,~~R(A_2)+R(B_2)+R(\phi_1)=2, \nonumber\\
&R(A_1)+R(B_1)+R(\phi_2)=2,~~R(A_2)+R(B_2)+R(\phi_2)=2, \nonumber\\
&R(\phi_1)=R(\phi_2)={2\over k+1}. 
\end{align}
We can use symmetry or $a$ maximization to find the following $R$ charges: $R(A_1)=R(A_2)=R(B_1)=R(B_2)={k\over k+1},~R(\phi_1)=R(\phi_2)={2\over k+1}$. 
The $F$-term relations from the superpotential are:
\begin{align}
& {\partial W\over \partial A_1}=0:~~B_1\phi_1-\phi_2 B_1=0, \nonumber\\
&{\partial W\over \partial B_1}=0:~~\phi_1 A_1-A_1\phi_2=0, \nonumber\\
&{\partial W\over \partial A_2}=0:~~B_2\phi_1-\phi_2 B_2=0, \nonumber\\
&{\partial W\over \partial B_2}=0:~~\phi_1 A_2-A_2\phi_2=0, \nonumber\\
&{\partial W\over \partial \phi_1}=0:~~A_1B_1+A_2B_2-2\phi_1^k=0, \nonumber\\
&{\partial W\over \partial \phi_2}=0:~~B_1A_1+B_2A_2-2\phi_2^k=0. \nonumber\\
\end{align}
The scalar chiral ring of this theory (which is related to the holomorphic functions on $X$) is generated by the loops in the quiver subject to the above relations. 
The single trace scalar chiral operators are generated by the simple loops, such as $\text{Tr}(A_iB_j)$, and  
one can order them by their $U(1)_R$ charge.

Consider the singularity $X$ defined by the equation $f=z_0^2+z_1^2+z_2^2+z_3^{2k}$, the unique candidate $U(1)_R$ symmetry $\zeta^{'}$ has charge 
$({2k\over k+1},{2k\over k+1},{2k\over k+1}, {2\over k+1})$ which is identified as the field theory $U(1)_R$ charge. We can make a holomorphic change of coordinates 
to write $f$ as $f=U^2+V^2+(-W+Z^k)(W+Z^k)$. 
The holomorphic functions on $X$ can be 
identified with the field theory chiral operators as follows:
\begin{align}
 \text{Tr}A_1B_2&=U,\qquad &\text{Tr}A_2 B_1&=V, \nonumber\\
 \text{Tr}A_1B_1&=-W+Z^k,\qquad &\text{Tr}A_2B_2&=W+Z^k. \nonumber\\
 \text{Tr}\phi_1&=Z. 
\end{align}
It can be checked that the full set of scalar chiral operators of the field theory which can get expectation value is captured by the ring $X$.

The properties of the IR SCFT can be derived as follows: The quiver without the superpotential term $\text{Tr}(\phi_1^{k+1})$ and $\text{Tr}(\phi_2^{k+1})$ 
defines a four dimensional $\mathcal{N}=2$  SCFT ${\cal T}_0$, and all of the elementary fields $A_i, B_i, \phi_i$ are free with $U(1)_R$ charge ${2\over 3}$. 
Our $\mathcal{N}=1$ theory can be thought of as deforming $\mathcal{N}=2$ SCFT ${\cal T}_0$ by the 
superpotential terms involving the adjoint chiral superfields $\phi_1$ and $\phi_2$. 
The scaling dimensions for the superpotential terms ${\cal O}_1=\text{Tr}(\phi_1^{k+1})$ and ${\cal O}_2=\text{Tr}(\phi_2^{k+1})$ are  $\Delta[{\cal O}_i]=k+1$, and so they
are irrelevant for $k>2$; 
the IR SCFT is just the original SCFT ${\cal T}_0$. 
For $k=2$, the superpotential term is marginally irrelevant \cite{Green:2010da}, and the IR SCFT is also the original SCFT ${\cal T}_0$.

We have used K stability to check that the ring is unstable for $k\geq 2$, and the destabilizing configuration has a central fibre $X_0: f=z_0^2+z_1^2+z_2^2$. 
It is interesting to note that the IR SCFT  (affine $A_1$, $\mathcal{N}=2$ SCFT) associated with the ring $X$ is  actually described by $D3$ branes probing the singularity $X_0$. This fact supports our claim that the central fibre $X_0$ describes the possible chiral ring of the IR SCFT, and the result from K-stability is in agreement with field theory result!

\subsubsection{Further obstructions}
The unstable example considered so far have been caused by the irrelevance of superpotential terms or the violation of the unitarity bound. 
We now give an example where the instability of the chiral ring is more subtle.  Consider a singularity
\begin{equation}
f=z_0^2+z_1^2+z_2^p+z_2z_3^q. 
\end{equation}
The only possible $U(1)_R$ symmetry has charge $({pq\over p+q-1},{pq\over p+q-1},{2q\over p+q -1},{2(p-1)\over p+q-1})$. The scaling dimensions of $z_2$ and $z_3$ are
\begin{equation}
[z_2]={3q\over 2(p+q-1)},~~[z_3]={3}{(p-1)\over p+q-1}.
\end{equation}
using the relation $\Delta({\cal O})={3\over2}R({\cal O})$.
The unitarity bound on the scalar operators implies that $[z_2]>1$ and $[z_3]>1$, we find 
\begin{equation}
p<2q+1~\&~q<2p-2.
\label{uni}
\end{equation}
The unitarity bound can also be found using the test configuration generated by the symmetry acting on coordinate $z_2$ and $z_3$ only. 

Consider a test configuration generated by the symmetry $\eta$ with charge $(0,0, 1, -1/q)$.  We have the following family generated by $\eta$;
\begin{equation}
z_0^2+z_1^2+t^pz_2^p+z_2z_3^q=0.
\end{equation}
The flat limit over $t=0$ is described by the equation $z_0^2+z_1^2+z_2 z_3^q=0$. The Futaki invariant can be computed using the formula \ref{hyper}:
\begin{equation}
F(X_0, \zeta, \eta)=-\frac{(p+q-1)^2 \left(p^2-2 p q+q-1\right)}{2 (p-1)^2 q^2}.
\end{equation}
So the original ring is stable if 
\begin{equation}
q(2p-1)-(p^2-1)>0 \rightarrow q>{p^2-1\over 2p-1}. 
\label{stronger}
\end{equation}
This bound is stronger than the unitarity bound~\eqref{uni} for certain range of the parameters. Let's set $p=6$, the unitarity bound from~\eqref{uni} implies that 
\begin{equation}
   {5\over 2}<q<10.
\end{equation}  
The bound from~\eqref{stronger} implies that 
\begin{equation}
q>34/11,
\end{equation}
which gives a stronger lower bound, i.e. $q=3$ satisfies the unitarity bound, but is unstable due to some other dynamical reason. The chiral ring of the IR SCFT is 
described by $z_0^2+z_1^2+z_2 z_3^q=0$ for $q=3$ which is also a three dimensional quotient singularity. 

\section{Conclusion}
We introduce a notion of stability for $\mathcal{N}=1$ chiral rings, and conjecture that a chiral ring is the chiral ring of a SCFT if and only if 
it is stable. We test our stability notion for models engineered using D3 brane probing 3-fold singularity, and show
that the notion of K-stability for the existence of Ricci-flat conic metric is equivalent to the field theory stability. This notion can be used to explain $a$-maximization, an operator becoming free if 
it violates unitarily bound, and the irrelevance of superpotential terms, etc. In general, our stability notion explains the consequences of accidental symmetries appearing in the study of quiver gauge theory: the chiral ring of the IR SCFT is different form that 
of UV theory if there are accidental symmetries.
Accidental symmetries cause many problems in studying supersymmetric field theory with four supercharges \cite{Kutasov:2003iy,Buican:2011ty}. Our study shows the importance 
of the chiral ring, and shows that the generalized notion of $a$-maximization plays a key role. Similar notion of generalized a maximization idea has already been used by Intriligator 
to settle some interesting IR phase questions \cite{Intriligator:2005if}. It would be interesting to use our stability notion to reconsider those models. 

The stability notion proposed here can be generalized to three dimensional $\mathcal{N}=2$ theory.  Although one does not have the central charge notion in this context, we may replace it by the so-called $F$-function \cite{Jafferis:2010un}. For the theory engineered by M2 branes probing a four-fold singularity, one still has the notion of K-stability for the four-fold singularity and much of the theory is similar. We leave the details to the interested reader. 
Similarly, one can also define the a notion of stability for two dimensional $(0,2)$ theory,  
and we hope that the accidental symmetry for $(0,2)$ theory studied in  \cite{Bertolini:2014ela} can be put into the stability framework.

There are some further questions about the stability of $\mathcal{N}=1$ chiral ring.
Of crucial importance is to understand the constraints on set of possible  test chiral rings.   At present, unless a large symmetry group intervenes, there are infinite number of possible test rings, and it seems computational impossible to check all of them, even in basic examples. It would be nice to have some physical input which could shed some light on this issue. In this paper, we only studied the models engineered using D3 brane, and it will be of great interest to study other $\mathcal{N}=1$ theories. 

Our primary focus has been on testing whether a chiral ring is the chiral ring of a SCFT. If the chiral ring is unstable, it is important to determine the ring of IR SCFT. Our study shows that the central fibre of the destabilizing test configuration should be the candidate chiral ring of the IR SCFT, and it is interesting to 
determine the special destabilizing test configuration which would give the chiral ring of IR SCFT. We hope to come to this question in the future.

%%%%%%%%%%%%%%%%%%%%%%%%%%%%%%%%%%%%  acknowledgements
\section*{Acknowledgments}
The work of S.T Yau is supported by  NSF grant  DMS-1159412, NSF grant PHY-
0937443, and NSF grant DMS-0804454.  T.C. Collins is supported by NSF grant DMS-1506652
The work of D. Xie is supported by Center for Mathematical Sciences and Applications at Harvard University, and in part by the Fundamental Laws Initiative of
the Center for the Fundamental Laws of Nature, Harvard University.

\bibliographystyle{JHEP}
\bibliography{kstab1}

\end{document}